\documentclass[preprint,12pt]{elsarticle}

\usepackage{amsmath}
\usepackage{amssymb}
\usepackage{graphicx}
\usepackage{booktabs}
\usepackage{hyperref}
\usepackage{geometry}
\journal{International Journal of Heat and Mass Transfer}

\begin{document}
	
	\begin{frontmatter}
		
		\title{Analytical solution of the Langmuir model for moisture diffusion in cylindrical coordinates.}
		
		\author[1]{Corentin Guigot\corref{cor1}\fnref{fn1}}
		\author[1]{Romain Grangeat\fnref{fn1}}
		\author[2]{Gilles-Alexis Renaut\fnref{fn1}}
		
		\fntext[fn1]{These authors contributed equally to this work.}
		
		\address[1]{CESI LINEACT, Campus CESI, 24, Le Paquebot - CS 60133, St-Nazaire, 44600, France}
		\address[2]{CESI, Campus CESI, 24, Le Paquebot - CS 60133, St-Nazaire, 44600, France}
		
		\cortext[cor1]{Corresponding author. Email address: cguigot@cesi.fr (C. Guigot)}
		
		\begin{abstract}
			
			Moisture diffusion in polymers and bio-based materials frequently exhibits non-Fickian behavior that cannot be described by classical diffusion models. The Langmuir model, which accounts for the coexistence of mobile and bound water molecules, has been widely used to represent such phenomena. However, analytical solutions of this model are generally limited to planar geometries, while cylindrical systems are typically investigated using numerical methods.
			
			In this work, the Langmuir diffusion model is solved analytically in cylindrical coordinates. The resulting solution provides both the local evolution of moisture content within the cylinder and the corresponding global moisture uptake kinetics. The analytical solution is validated through comparison with an independent numerical solution based on a finite difference scheme, showing excellent agreement for both the global absorption kinetics and the radial moisture profiles.
			
			The proposed formulation therefore provides a simple and efficient analytical framework for studying non-Fickian moisture diffusion in cylindrical systems such as natural fibers, and facilitates the identification of model parameters from experimental data.
			
		\end{abstract}
		
		\begin{keyword}
			Moisture diffusion \sep Langmuir model \sep Cylindrical geometry \sep Analytical solution
		\end{keyword}
		
	\end{frontmatter}
	
	\section{Introduction}
	\label{sec:intro}
	
	Water diffusion in organic materials plays a key role in the durability and performance of polymers, composites and bio-based materials. Moisture absorption can indeed lead to significant changes in mechanical properties, dimensional stability and long-term reliability of these materials \cite{Bachchan2022}. Although water diffusion is often described using Fick’s law \cite{Crank1975}, many polymeric and composite systems exhibit absorption kinetics that deviate from classical Fickian behavior \cite{Weitsman2012}. These deviations, commonly referred to as non-Fickian behavior, are generally attributed to interactions between water molecules and the polymer matrix or to the presence of trapping sites within the material.
	
	In order to describe these phenomena, several models have been proposed in the literature. Besides approaches based on dual-Fick formulations \cite{Placette2012}, the Langmuir model introduced by Carter and Kibler \cite{Carter1978} distinguishes two populations of water molecules: mobile molecules diffusing through the material and molecules temporarily bound to the matrix. This model has been widely used to analyze moisture absorption kinetics in various polymer and composite systems \cite{Popineau2005,Melo2020,Yuan2022,Teixeira2019}. However, analytical solutions of the model are generally established for simple geometries, particularly planar sheets. When more complex geometries are considered, the governing equations are most often solved using numerical methods. For instance, Célino et al. applied the Langmuir model to moisture diffusion in natural fibers approximated as cylinders by solving the equations using a finite difference scheme \cite{Celino2013}. To the authors’ knowledge, no analytical solution of the Langmuir model in cylindrical geometry has been reported in the literature. Such a solution would nevertheless be of significant interest, as it would allow a more direct analysis of diffusion mechanisms and facilitate the identification of model parameters from experimental data, while avoiding the systematic use of numerical methods.
	
	The objective of this study is therefore to analytically solve the differential equations of the Langmuir diffusion model in cylindrical coordinates in order to obtain an analytical expression for both the absorption kinetics and the moisture content distribution within the material.
	
	\section{Mathematical Framework: The Cylindrical Langmuir Solution}
	\label{sec:theory}
	
	The Langmuir physical model assumes that moisture exists in two states within the polymer matrix: mobile molecules (concentration $n$) and bound molecules (concentration $N$). The transport is governed by Fickian diffusion of the mobile phase, coupled with a reversible first-order chemical reaction describing the trapping mechanism. 
	
	For an infinite cylinder of radius $a$, assuming purely radial diffusion and constant boundary conditions ($n(a,t) = n_\infty$, $N(a,t) = N_\infty$), the coupled partial differential equations are \cite{Crank1975}:
	
	\begin{align}
		\frac{D}{r}\frac{\partial}{\partial r}\left(r \frac{\partial n}{\partial r}\right) &= \frac{\partial n}{\partial t} + \frac{\partial N}{\partial t} \label{eq:coupled1} \\
		\frac{\partial N}{\partial t} &= \gamma n - \beta N \label{eq:coupled2}
	\end{align}
	where $D$ is the intrinsic diffusion coefficient of the mobile molecules, $\gamma$ is the probability per unit time that a mobile molecule becomes bound, and $\beta$ is the probability per unit time that a bound molecule is released.
	
	While the mathematical resolution of analogous coupled diffusion-reaction systems in cylindrical coordinates was historically explored for finite-bath dyeing processes \cite{Wilson1948}, its explicit derivation for the constant boundary conditions characteristic of standard hygrothermal aging remains largely absent from the polymer literature. By applying separation of variables, evaluating the inverse Laplace transform via Cauchy's residue theorem \cite{Carslaw1959}, and exploiting the orthogonality properties of Bessel functions of the first kind of order zero ($J_0$) \cite{Watson1944}, the exact formulation for the local concentration of the mobile phase $n(r,t)$ and the bound phase $N(r,t)$ can be established. 
	
	The normalized total local concentration $C(r,t) / C_\infty$, where $C_\infty = n_\infty (1 + \gamma/\beta)$ is the thermodynamic equilibrium concentration, explicitly describes the spatio-temporal evolution of the moisture front within the cylindrical cross-section. It is expressed as:
	
	\begin{equation}
		\frac{C(r,t)}{C_\infty} = 1 - \sum_{k=1}^{\infty} \frac{2}{ \alpha_k J_1(\alpha_k)} J_0\left(r \frac{\alpha_k}{a}\right) \left[ \frac{R_k^+ e^{-R_k^- t} - R_k^- e^{-R_k^+ t}}{R_k^+ - R_k^-} - \frac{D \left(\frac{\alpha_k}{a}\right)^2 \beta}{\beta + \gamma} \frac{e^{-R_k^- t} - e^{-R_k^+ t}}{R_k^+ - R_k^-} \right]
		\label{eq:local_concentration}
	\end{equation}
	
	In this solution, $\alpha_k$ are the positive roots of $J_0(\alpha_k) = 0$. The temporal eigenvalues $R_k^\pm$ governing the relaxation rates of the coupled system are defined by:
	
	\begin{equation}
		R_k^\pm = \frac{1}{2} \left[ \left(D\left(\frac{\alpha_k}{a}\right)^2 + \gamma + \beta\right) \pm \sqrt{\left(D\left(\frac{\alpha_k}{a}\right)^2 + \gamma + \beta\right)^2 - 4 D\left(\frac{\alpha_k}{a}\right)^2 \beta} \right]
		\label{eq:eigenvalues}
	\end{equation}
	
	The physical validity of the model requires these temporal eigenvalues to be strictly real, forbidding any unphysical oscillatory behavior in the concentration history. This is verified by analyzing the discriminant $\Delta_k$ under the square root in Equation \ref{eq:eigenvalues}. By setting $\lambda_k = D(\alpha_k/a)^2$, the discriminant can be algebraically rearranged as:
	
	\begin{equation}
		\Delta_k = (\lambda_k - \beta)^2 + \gamma^2 + 2\gamma(\lambda_k + \beta)
	\end{equation}
	
	Since $D$, $\gamma$, and $\beta$ are strictly positive physical parameters, it follows that $\Delta_k > 0$ for all $k$. Consequently, the relaxation rates $R_k^\pm$ are always real, distinct, and positive, ensuring a purely dissipative transport process.

	Regarding the overall convergence of the analytical solution, Equation \ref{eq:local_concentration} behaves as a standard Fourier-Bessel expansion. For any macroscopic time $t > 0$, the temporal eigenvalues $R_k^\pm$ scale quadratically with the roots $\alpha_k$. Consequently, the exponential relaxation terms $e^{-R_k^\pm t}$ act as powerful regularizers, ensuring the absolute and uniform convergence of the infinite series over the entire closed spatial domain $r \in [0, a]$. The most critical point of evaluation remains the center of the cylinder ($r=0$), where $J_0(0) = 1$ and the spatial attenuation is null. Even at this singularity, the temporal exponential decay strictly suppresses the higher-order terms, thereby guaranteeing the mathematical stability, continuity, and physical validity of the concentration fields everywhere within the cylinder.
	
	Finally, by integrating the local concentration field over the cylindrical volume, the exact analytical solution for the normalized macroscopic mass uptake $M(t)/M_\infty$ is derived:
	
	\begin{equation}
		\frac{M(t)}{M_\infty} = 1 - \sum_{k=1}^{\infty} \frac{4}{\alpha_k^2} \left[ \frac{R_k^+ e^{-R_k^- t} - R_k^- e^{-R_k^+ t}}{R_k^+ - R_k^-} - \frac{D \left(\frac{\alpha_k}{a}\right)^2 \beta}{\beta + \gamma} \frac{e^{-R_k^- t} - e^{-R_k^+ t}}{R_k^+ - R_k^-} \right]
		\label{eq:mass_uptake_cyl}
	\end{equation}
	The rigorous mathematical derivation of these equations, from the Laplace domain to the time-domain inversion via Cauchy's residue theorem, is detailed in Appendix A.

	\section{Model validation}
	\label{sec:results}
	
\begin{figure}[htbp]
		\centering
		\includegraphics[width=0.85\linewidth]{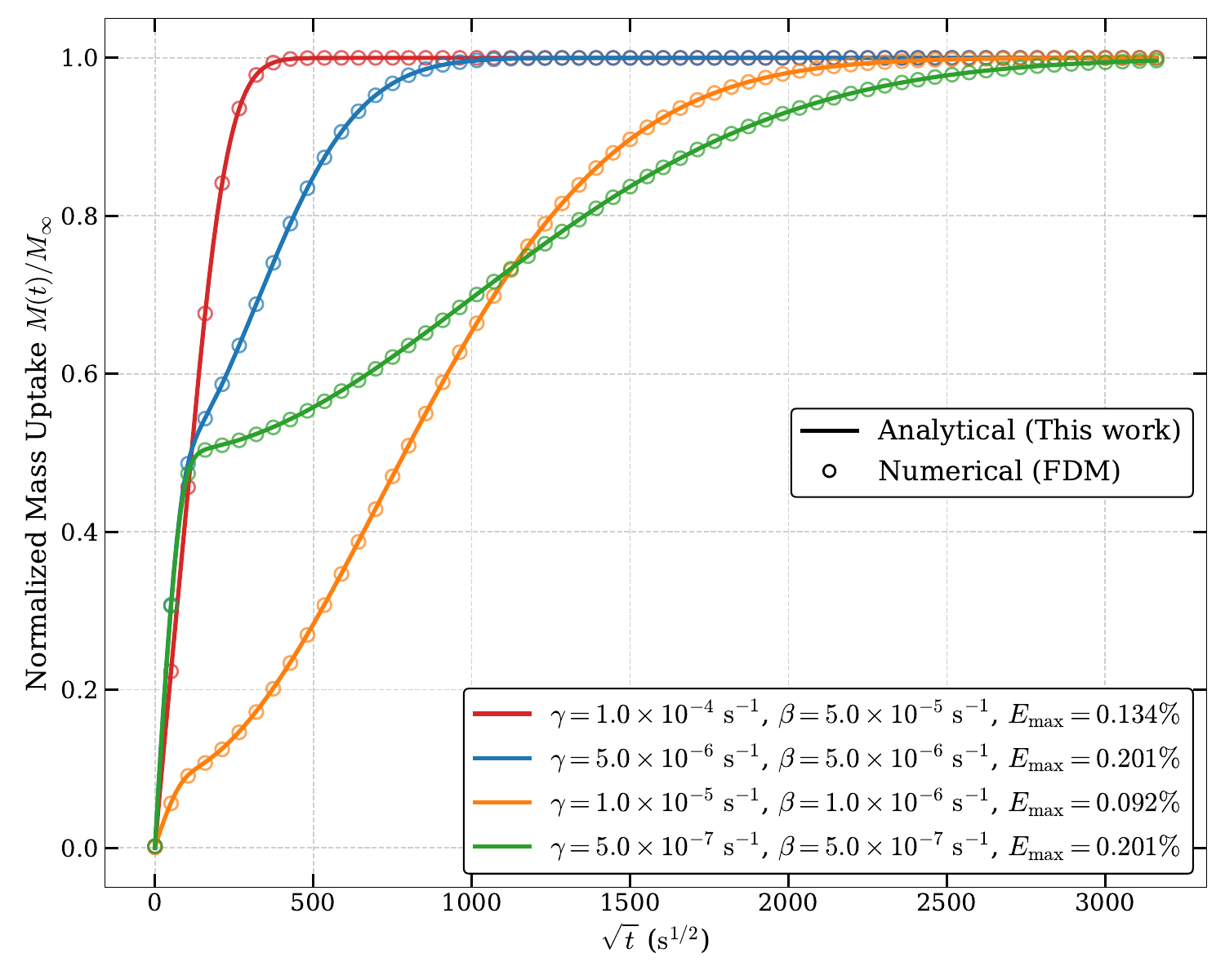}
		\caption{Comparison between the analytical solution and the numerical solution (finite difference method) for the evolution of the normalized moisture uptake for different Langmuir model parameters.}
		\label{fig:global}
	\end{figure}

	To verify the validity of the analytical solution, the results are compared with those obtained from an independent numerical resolution of the Langmuir model using the finite difference method. Figure~\ref{fig:global} presents the evolution of the normalized moisture uptake $M(t)/M_{\infty}$ obtained from both the analytical solution and the numerical solution for different sets of Langmuir model parameters. An excellent agreement between the two approaches is observed for all the cases studied (the maximum relative error is 0.201\%).
	
\begin{figure}[htbp]
		\centering
		\includegraphics[width=0.85\linewidth]{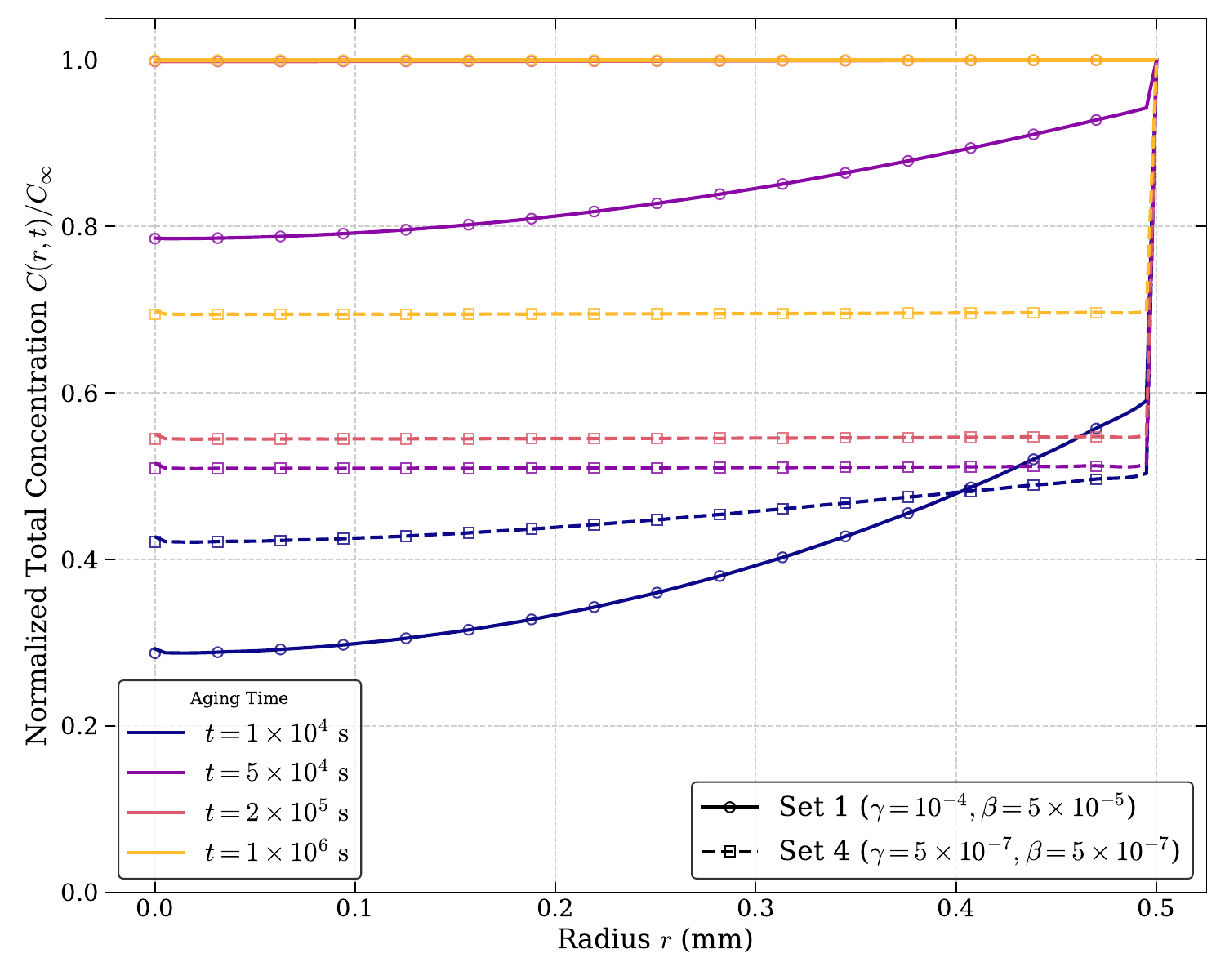}
		\caption{Comparison between the analytical solution and the numerical solution for the radial moisture content profiles at different times for two sets of parameters.}
		\label{fig:local}
	\end{figure}

	To further verify the accuracy of the solution at the local level, the radial distributions of moisture content are compared at different times. Figure~\ref{fig:local} shows the concentration profiles obtained from the analytical solution and the numerical resolution. Both approaches lead to nearly identical distributions over the entire radial domain and for all considered times, confirming the validity of the proposed analytical solution.
	
	\section{Conclusion}
	\label{sec:conclusion}
	
	In this work, an analytical solution of the Langmuir diffusion model in cylindrical geometry has been established. Solving the differential equations governing moisture transport makes it possible to determine both the local evolution of the moisture content fields within the cylinder and the corresponding global absorption kinetics.
	
	The validity of the analytical solution has been verified through comparison with a numerical resolution of the model based on the finite difference method.
	
	The proposed solution therefore provides a simple and efficient analytical tool for analyzing the Langmuir diffusion model in cylindrical geometry. It also opens interesting perspectives for modeling moisture diffusion in cylindrical systems such as natural fibers and enables rapid identification of the model parameters from experimental sorption curves.

	\appendix
	\section{Rigorous Derivation of the Cylindrical Langmuir Solution}
	\label{app:derivation}
	
	This section details the exact analytical resolution of the Langmuir-type coupled partial differential equations in cylindrical coordinates using the Laplace transform and Cauchy's residue theorem.
	
	\subsection{Governing Equations and Boundary Conditions}
	Consider an infinite solid polymer cylinder of radius $a$. Moisture is assumed to diffuse purely in the radial direction $r$. The Langmuir physical model distinguishes mobile water molecules of concentration $n(r,t)$ and bound water molecules of concentration $N(r,t)$. The coupled transport-trapping system is governed by:
	\begin{align}
		\frac{D}{r}\frac{\partial}{\partial r}\left(r \frac{\partial n}{\partial r}\right) &= \frac{\partial n}{\partial t} + \frac{\partial N}{\partial t} \label{eq:a1} \\
		\frac{\partial N}{\partial t} &= \gamma n - \beta N \label{eq:a2}
	\end{align}
	where $D$ is the diffusion coefficient, $\gamma$ is the trapping probability, and $\beta$ is the release probability. 
	The specimen is initially dry: $n(r,0) = 0$ and $N(r,0) = 0$. Upon immersion, a constant surface concentration of the mobile phase is instantaneously reached and maintained: $n(a,t) = n_\infty$.
	
	\subsection{Laplace Transform}
	Applying the Laplace transform with respect to time ($t \to p$) to the initial system yields:
	\begin{align}
		\frac{D}{r}\frac{d}{dr}\left(r \frac{d\bar{n}}{dr}\right) &= p\bar{n} + p\bar{N} \label{eq:a3} \\
		p\bar{N} &= \gamma \bar{n} - \beta \bar{N} \label{eq:a4}
	\end{align}
	Isolating $\bar{N}$ from Eq. \ref{eq:a4} gives the direct relation in the Laplace domain: 
	\begin{equation}
		\bar{N}(r,p) = \frac{\gamma}{p+\beta}\bar{n}(r,p)
		\label{eq:a5}
	\end{equation}
	Substituting this relation into Eq. \ref{eq:a3} leads to the modified Bessel differential equation of order zero for the mobile phase:
	\begin{equation}
		\frac{d^2\bar{n}}{dr^2} + \frac{1}{r}\frac{d\bar{n}}{dr} - \mu^2 \bar{n} = 0 \quad \text{with} \quad \mu^2 = \frac{p}{D}\left(\frac{p+\gamma+\beta}{p+\beta}\right)
	\end{equation}
	The general solution involves modified Bessel functions of the first ($I_0$) and second ($K_0$) kind. To maintain a finite concentration at the center of the cylinder ($r=0$), the coefficient associated with $K_0(\mu r)$ must be strictly zero. Applying the transformed boundary condition $\bar{n}(a,p) = n_\infty/p$, the spatial distribution of the mobile phase is:
	\begin{equation}
		\bar{n}(r,p) = \frac{n_\infty}{p} \frac{I_0(\mu r)}{I_0(\mu a)}
	\end{equation}
	Injecting this result back into Eq. \ref{eq:a5} yields the bound phase distribution. The total local concentration $\bar{C}(r,p) = \bar{n}(r,p) + \bar{N}(r,p)$ is therefore obtained as:
	\begin{equation}
		\bar{C}(r,p) = \frac{n_\infty}{p} \left( \frac{p+\gamma+\beta}{p+\beta} \right) \frac{I_0(\mu r)}{I_0(\mu a)}
	\end{equation}
	
	\subsection{Inverse Laplace Transform and Residue Theorem}
	The time-domain local concentration $C(r,t)$ is recovered using the Mellin-Fourier complex inversion integral, evaluated via Cauchy's residue theorem over the poles of $\bar{C}(r,p)$.
	
	At the simple pole $p=0$, $\mu^2 = 0$, implying $I_0(0) = 1$. The residue defines the thermodynamic saturation concentration $C_\infty$:
	\begin{equation}
		\text{Res}(p=0) = n_\infty \left( \frac{\gamma+\beta}{\beta} \right) = C_\infty
	\end{equation}
	
	The transient diffusion dynamics are dictated by the infinite number of roots of the denominator $I_0(\mu a) = 0$. Using the identity $I_0(ix) = J_0(x)$, where $J_0$ is the Bessel function of the first kind of order zero, the roots occur when $\mu a = i\alpha_k$, with $\alpha_k$ being the strictly positive roots of $J_0(\alpha_k) = 0$. This structural condition imposes $\mu^2 = -(\alpha_k/a)^2$. Substituting the definition of $\mu^2$:
	\begin{equation}
		\frac{p}{D}\left(\frac{p+\gamma+\beta}{p+\beta}\right) = -\left(\frac{\alpha_k}{a}\right)^2
	\end{equation}
	Introducing the cylindrical spatial eigenvalue $\lambda_k = D(\alpha_k/a)^2$, we obtain the characteristic quadratic equation of the Langmuir system:
	\begin{equation}
		p^2 + p(\lambda_k + \gamma + \beta) + \lambda_k \beta = 0
	\end{equation}
	This equation yields two sets of strictly real and negative roots $p_k^\pm$. To ensure exponential decay in the time domain, we define the strictly positive relaxation rates $R_k^\pm = -p_k^\mp$. The roots $R_k^\pm$ of the characteristic equation $R^2 - R(\lambda_k + \gamma + \beta) + \lambda_k \beta = 0$ are expressed as:
	\begin{equation}
		R_k^\pm = \frac{1}{2}\left[ (\lambda_k + \gamma + \beta) \pm \sqrt{(\lambda_k + \gamma + \beta)^2 - 4\lambda_k \beta} \right]
	\end{equation}
	
	Evaluating the residues at the simple poles $p_k \in \{p_k^+, p_k^-\}$ requires computing the complex derivative of the denominator $D(p) = p I_0(\mu a)$ with respect to $p$. At the poles, $I_0(\mu a) = 0$, hence $D'(p_k) = p_k \left. \frac{d}{dp}[I_0(\mu a)] \right|_{p=p_k}$. Applying the chain rule and the identity $I_1(i\alpha_k) = i J_1(\alpha_k)$ yields:
	\begin{equation}
		\left. \frac{d}{dp}[I_0(\mu a)] \right|_{p=p_k} = a I_1(i\alpha_k) \left. \frac{d\mu}{dp} \right|_{p=p_k} = \frac{a^2 J_1(\alpha_k)}{2 \alpha_k} \left. \frac{d(\mu^2)}{dp} \right|_{p=p_k}
	\end{equation}
	Differentiating $\mu^2(p)$ analytically gives $\frac{d(\mu^2)}{dp} = \frac{1}{D} \left( 1 + \frac{\gamma\beta}{(p+\beta)^2} \right)$. Consequently, the exact derivative of the denominator is:
	\begin{equation}
		D'(p_k) = p_k \frac{a^2 J_1(\alpha_k)}{2 D \alpha_k} \left( 1 + \frac{\gamma\beta}{(p_k+\beta)^2} \right)
	\end{equation}
	Combining this derivative with the numerator $N(p_k) = n_\infty \left( \frac{p_k+\gamma+\beta}{p_k+\beta} \right) J_0(r\alpha_k/a)$ and summing the contributions of the paired poles $p_k^+$ and $p_k^-$ simplifies the algebraic fractions. Expanding $\lambda_k$ back to $D(\alpha_k/a)^2$ yields the exact local concentration profile:
	\begin{equation}
		\frac{C(r,t)}{C_\infty} =  1 - \sum_{k=1}^{\infty} \frac{2 J_0(r \alpha_k/a)}{\alpha_k J_1(\alpha_k)} \left[ \frac{R_k^+ e^{-R_k^- t} - R_k^- e^{-R_k^+ t}}{R_k^+ - R_k^-} - \frac{D\left(\frac{\alpha_k}{a}\right)^2 \beta}{\gamma+\beta} \frac{e^{-R_k^- t} - e^{-R_k^+ t}}{R_k^+ - R_k^-} \right] 
	\end{equation}
	
	\subsection{Macroscopic Mass Uptake Integration}
	The macroscopic mass uptake $M(t)$ is obtained by integrating the local concentration field over the circular cross-section of the cylinder:
	\begin{equation}
		M(t) = \int_0^a 2\pi r \, C(r,t) \, dr
	\end{equation}
	The spatial integration of the Bessel function strictly adheres to the fundamental property $\int_0^a r J_0(r\alpha_k/a) dr = \frac{a^2}{\alpha_k} J_1(\alpha_k)$. Applying this integral operator transforms the geometric pre-factor $\frac{2}{\alpha_k J_1(\alpha_k)}$ perfectly into $\frac{4}{\alpha_k^2}$. Normalizing by the saturation mass $M_\infty = \pi a^2 C_\infty$ yields the final, exact analytical solution:
	\begin{equation}
		\frac{M(t)}{M_\infty} = 1 - \sum_{k=1}^{\infty} \frac{4}{\alpha_k^2} \left[ \frac{R_k^+ e^{-R_k^- t} - R_k^- e^{-R_k^+ t}}{R_k^+ - R_k^-} - \frac{D \left(\frac{\alpha_k}{a}\right)^2 \beta}{\beta + \gamma} \frac{e^{-R_k^- t} - e^{-R_k^+ t}}{R_k^+ - R_k^-} \right]
	\end{equation}

	\bibliographystyle{elsarticle-num}
	\bibliography{letter_bib}
	
\end{document}